\documentclass[sigconf]{acmart}
\usepackage[nolist,nohyperlinks]{acronym}
\usepackage{multirow}
\usepackage{bbm}
\usepackage{amsmath}
\usepackage{hyperref}
\usepackage{subfig}
\AtBeginDocument{%
  }

\copyrightyear{2025}
\acmYear{2025}
\setcopyright{acmlicensed}\acmConference[UMAP '25]{33rd ACM Conference on User Modeling, Adaptation and Personalization}{June 16--19, 2025}{New York City, NY, USA}
\acmBooktitle{33rd ACM Conference on User Modeling, Adaptation and Personalization (UMAP '25), June 16--19, 2025, New York City, NY, USA}
\acmDOI{10.1145/3699682.3728331}
\acmISBN{979-8-4007-1313-2/2025/06}

\begin{document}

\title{Uncertainty in Repeated Implicit Feedback as a Measure of Reliability}

\author{Bruno Sguerra}
\authornote{Contact author: \href{research@deezer.com}{research@deezer.com}}
\affiliation{
  \institution{Deezer Research}    
  \country{France}
}

\author{Viet-Anh Tran}
\affiliation{
  \institution{Deezer Research}
  \country{France}
}

\author{Romain Hennequin}
\affiliation{
  \institution{Deezer Research}
    \country{France}
}
\author{Manuel Moussallam}
\affiliation{
  \institution{Deezer Research}
    \country{France}
}

\renewcommand{\shortauthors}{Sguerra et al.}

\begin{abstract}
Recommender systems rely heavily on user feedback to learn effective user and item representations. Despite their widespread adoption, limited attention has been given to the uncertainty inherent in the feedback used to train these systems. Both implicit and explicit feedback are prone to noise due to the variability in human interactions, with implicit feedback being particularly challenging. In collaborative filtering, the reliability of interaction signals is critical, as these signals determine user and item similarities. Thus, deriving accurate confidence measures from implicit feedback is essential for ensuring the reliability of these signals.

A common assumption in academia and industry is that repeated interactions indicate stronger user interest, increasing confidence in preference estimates. However, in domains such as music streaming, repeated consumption can shift user preferences over time due to factors like satiation and exposure. While literature on repeated consumption acknowledges these dynamics, they are often overlooked when deriving confidence scores for implicit feedback.

This paper addresses this gap by focusing on music streaming, where repeated interactions are frequent and quantifiable. We analyze how repetition patterns intersect with key factors influencing user interest and develop methods to quantify the associated uncertainty. These uncertainty measures are then integrated as consistency metrics in a recommendation task. Our empirical results show that incorporating uncertainty into user preference models yields more accurate and relevant recommendations. Key contributions include a comprehensive analysis of uncertainty in repeated consumption patterns, the release of a novel dataset, and a Bayesian model for implicit listening feedback.

\end{abstract}

\begin{acronym}
    \acro{ALS}{Implicit Matrix Factorization}
    \acro{MF}{Matrix Factorization}
    \acro{DCG}{Discounted Cumulative Gain}
    \acro{NDCG}{Normalized Discounted Cumulative Gain}
    \acro{CF}{Collaborative Filtering}
    \acro{ACT-R}{Adaptive Control of Thought—Rational}
    \acro{SVD}{Singular Value Decomposition}
    \acro{GNN}{graph neural network}
    \acro{NeuMF}{Neural collaborative filtering}
    \acro{WNeuMF}{Weighted Neural Collaborative Filtering}
    \acro{MEE}{Mere Exposure Effect}
    \acro{LE}{Listening Event}
    \acro{BLA}{Base-Level Activation}
    \acro{UQ}{Uncertainty Quantification}
    \acro{NN}{Neural Networks}
    \acro{pdf}{probability density function}
    \acro{HDI}{High Density Interval}
    \acro{AU}{Aleatoric Uncertainty}
    \acro{EU}{Epistemic Uncertainty}
    
\end{acronym}


\maketitle

\section{Introduction}
``Uncertainty is an uncomfortable position. But certainty is an absurd one'' is a quote credited to the French philosopher Voltaire. Voltaire understood that, while dealing with uncertain situations is uncomfortable, it is necessary, as certainty can restrict self-reflection and questioning. While people can manage everyday life's uncertain decisions, within some situations with predictable errors~\cite{tversky1978judgment}, recommender systems are plagued with a rather specific type of uncertainty, one derived from noisy user feedback. 

In broad terms, \ac{CF}-based recommender systems are divided into two main categories, given the type of information used to model user preferences: explicit and implicit feedback~\cite{jawaheer2014modeling}. Explicit feedback is information users directly provide about their preferences for items, such as movie ratings or thumbs-up/down indicators. In contrast, implicit feedback encompasses all collections of actions and behavior not explicitly meant to provide feedback on items or system performance. In this case, the feedback is derived from user interactions, such as the number of clicks on an item or the time spent on the platform. Both categories of feedback are subject to uncertainty derived from the natural noise of human interactions. In particular, implicit feedback is arguably noisier than its explicit counterpart. Users' interactions traits with recommender systems have many different causes than pure interest. For example, a user might interact with an item out of curiosity rather than preferences or might leave their session open while doing something else, increasing dwell time. There are a multitude of reasons to interact with different type of content besides pure preferences. 

In the often employed \ac{CF} setting for recommendation, it is very common to compare interaction signals (or lack thereof) between items across users to derive preference estimations~\cite{hu2008collaborative, ferraro2020maximizing}. For these interaction signals to be meaningful, the signal between users and items needs to be reliable for deriving accurate similarity measures, which in turn drive the algorithm’s training. By \textit{reliable}, we mean that the interaction signals should be clear indicators of the same underlying factor—user interest. In the context of implicit feedback, due to its noisiness, it becomes essential to derive measures of confidence to ensure that the interaction signals accurately represent user preferences. A commonly employed premise is that the higher the number of repeated interactions a user has with an item, the more confident one can be about the user's interest. However, in some media consumption domains such as music streaming (also books, movies, and food consumption), repeated consumption of items affects user's perceptions of the items and in consequence their preferences. Notably, the literature on repeated consumption highlights factors like satiation and multiple exposures as key modulators of user interest~\cite{anderson2014dynamics, benson2016modeling, sguerra2022discovery, sguerra2023ex2vec}. Here we posit that patterns of repeated consumption should be leveraged to derive confidence scores, as they provide valuable information characterizing users' interest evolution dynamics.

In this paper, we propose to tackle this important challenge. We turn our focus to music streaming consumption where the effects of repeated consumption are easily observable from data, as music repetition is rather common~\cite{conrad2019extreme,tsukuda2020explainable}. 
In Section~\ref{sec:uncertainty}, we first present the reader with an in-depth exploration of uncertainty quantification in real music streaming data over different factors that are known to modulate users' interest. To the best of our knowledge, this is the first uncertainty exploration of implicit feedback. We empirically show that uncertainty dynamics correlate with known repetitive patterns. In Section~\ref{sec:reliability} we expand on how to derive measures of uncertainty for any user interaction with an item and how they relate to feedback reliability. Section~\ref{sec:experiments} reports our experiments in testing our uncertainty measures as confidence scores in implicit feedback for a recommendation task. Lastly, Section~\ref{sec:conclusion} discusses our findings and future directions. 

More specifically, our contributions in this paper are listed as follows:
\begin{itemize}
    \item we introduce a Bayesian modeling of implicit listening feedback;
    \item we perform an in-depth exploration of uncertainty patterns over repeated music consumption;
    \item we provide empirical evidence that our approach improves preference modeling in a recommendation task;
    \item we release the code of our experiments to ensure the reproducibility of our results and facilitate its future use. Additionally, we release our dataset of listening consumption.
\end{itemize}

\section{Related Work}
This work is related to two different areas, reliability in implicit feedback and repeated consumption, both discussed here.

\textbf{Reliability in implicit feedback}.
\label{sec:related_work}
Given the inherent noisiness of human interactions, feedback reliability, or the degree to which feedback accurately captures true preferences, has been the subject of extensive research. However, reliability in implicit feedback has been relatively less explored compared to explicit feedback. In explicit settings, users ratings are often inconsistent due to changes in attention, motivation, context, etc. A common technique to quantify this inconsistency is to ask users to re-rate items at different times, often resulting in disagreements~\cite{jawaheer2014modeling}.  An example of this approach is presented in~\cite{amatriain2009rate}, where Amatriain et al. researched the effects of natural noise in movie ratings. Users were asked to rate the same movies at three different times in varying orders, creating three datasets for agreement analysis in order to identify and remove noise. They show that, due to the denoising, the recommendation accuracy increased considerably.  Another method of denoising explicit feedback was proposed in~\cite{o2006detecting}, where the authors propose a threshold to limit deviations between user ratings and predicted ratings. They demonstrate that removing ratings exceeding this threshold improves recommendation accuracy.

However, implicit feedback is arguably a noisier signal of user preference than the explicit counterpart. Clicks and browsing time are harder to interpret than the feedback the user actually intends to give the system. Note that here we focus on the context of collaborative filtering which are the basis for most recommender systems~\cite{cremonesi2010performance}. One popular strategy to mitigate this noisiness is to leverage multiple consumption of the same item. Hu et al., in~\cite{hu2008collaborative}, propose a confidence score. The score, denoted $c_{u,i}$, represents the confidence of observing the interest of a pair user $u$ and item $i$ based on the total number of interactions $r_{u,i}$. They propose two formulations for $c_{u,i}$: (i) a linear one, $c_{u,i} = 1 + \alpha \, r_{u,i}$ and (ii) a logarithmic one, $c_{u,i} = 1 + \alpha \, log(1 + r_{u,i}/\epsilon)$, controlled by the constants $\alpha$ and $\epsilon$. The premise is that \textit{``...as $r_{ui}$ grows, we have a stronger indication that the user indeed likes the item''}. The confidence interval is used in the cost function of their \ac{ALS} model. More recently, authors in~\cite{hennekes2023weighted} extend on the confidence scores proposed in~\cite{hu2008collaborative}. First, they account for the average number of interactions of item $i$, noted $\tilde{r_i}$, defining the confidence of the positive interaction as $c_{u,i} = 1 + \alpha \, log(1 + r_{u,i}/(\tilde{r_i} \epsilon))$. Second, inspired by~\cite{he2016fast}, they define a weight to the non-interacted items based on item popularity, the rational being that popular items that were not interacted with have more chance of being known beforehand, and in consequence better negative samples. And finally, they broaden the model beyond \ac{ALS} by extending the popular \ac{NeuMF} model~\cite{he2017neural} by incorporating the positive weights in the loss function and the negative ones for negative sampling, their model is named \ac{WNeuMF}. 

Another strategy, more close, in principle, to~\cite{o2006detecting}, consists in adjusting (often removing) training data that is far from what is expected. Wang et al., in~\cite{wang2021denoising}, observe that noisy observations have larger loss values, more so on the first stages of training. In light of this observation, they introduce two novel adaptive loss strategies: (i) Truncated Loss, which dynamically discards high-loss samples using a threshold tailored for each iteration, and (ii) Reweighted Loss, which dynamically adjusts the weights of high-loss samples to mitigate their impact on training.  Using the same principle,~\cite{tian2022learning} design a \ac{GNN} method for reducing the impact of noisy interactions on the representation learning of \ac{GNN}. Inspired by the homophily theory (stating that individuals tend to form connections with others who are similar to themselves) they derive a reliability score from the structural similarity of a graph derived from user-item interaction data. They go on to use this score to either filter out low reliability interactions or assign a weight to the remaining ones.

\textbf{Repeated consumption}. Repeated consumption is well documented across various domains such as grocery shopping, business visits, and music listening. Benson et al.~\cite{benson2016modeling} define it as a process wherein users discover new items, consume them frequently, and then shift their attention to other alternatives. This pattern often leads to a widening gap between repeated consumption instances over time, interpreted as growing boredom or satiation~\cite{benson2016modeling, anderson2014dynamics}. Other than satiation, in our previous work, we expand on the~\ac{MEE} and its implications for recommender systems in music streaming platforms~\cite{sguerra2022discovery,sguerra2023ex2vec}. We show that, over repeated exposures to newly released songs, users' interest tends to follow a well-documented inverted-U shape~\cite{bornstein1989exposure,montoya2017re}. That is, aggregated listening histories reveal that interest typically starts low, increases as users become more familiar with the songs, and then gradually declines as users grow satiated.

Repeated consumption holds significant relevance for recommender system development, as users exhibit repeated interactions with items across diverse domains~\cite{tsukuda2020explainable}. Several studies have explored repetition and its influencing factors, such as boredom, in the design of new recommender frameworks e.g.~\cite{hu2011nextone,rafailidis2015repeat,chmiel2018simple,ren2019repeatnet}. In recent years, there has been a growing popularity of using Anderson's cognitive architecture \ac{ACT-R}~\cite{anderson2009can}, particularly its declarative module, for modeling user repeated consumption ~\cite{kowald2017temporal, knees2019user,reiter2021predicting, moscati2023integrating, sguerra2023ex2vec, tran2024transformers}. ACT-R's declarative module models the dynamics of human memory and it has terms accounting for both recency (how recent is the interaction) and frequency (how often the item was interacted with), which is rather effective in modeling user repeat behavior.

As previously mentioned, both~\cite{hu2008collaborative, he2016fast, hennekes2023weighted} increase the level of confidence in function of the number of observed interactions based on a common premise: the higher the number of user interactions with an item, the more certain about the user's preferences for the item. This is a common premise, found in both academia and industry, as user feedback is typically measured as some form of counting of repeatable behaviors~\cite{jawaheer2014modeling}, if not simply binarized~\cite{jannach2018recommending}. For instance, the number of repeated interactions with a song is the main engagement  signal used for measuring how much a user likes an artist~\cite{ferraro2020maximizing}. Incidentally the amount of ``plays'' of a song is also considered as the primary indicator of its success among the public, having replaced the sales numbers despite carrying a somewhat different meaning.

While the assumption above seems intuitive, it has seldom been tested. A higher number of interactions with an item does not necessarily characterize more reliable feedback. Moreover it also overlooks the dynamics of repeated consumption: the more users reconsume the more satiated they get. In addition, accounting for the~\ac{MEE}, the relationship between interest and interaction number becomes quadratic (inverted-U shape). In the upcoming sessions, we aim to fill the gap between these two areas by deriving confidence estimates from repeated consumption.

\section{Uncertainty in Musical implicit feedback}
\label{sec:uncertainty}
As emphasized in the previous section, phenomena like satiation and the~\ac{MEE} significantly affect users' musical preferences over repeated interactions.  In~\cite{sguerra2022discovery}, it is shown that the interest curve for a song typically requires several exposures before it reaches its peak. According to the literature on the~\ac{MEE}, the number of exposures required for liking to peak varies depending on the specific stimuli involved ~\cite{bornstein1989exposure, montoya2017re}. Therefore, when assessing interest through the amount of play counts, a low number of interactions might indicate a lack of interest, but also, it might suggest that the interest hasn't yet reached its peak. Distinguishing these cases is important when modeling preferences.
Additionally, users often like/rate differently the same songs at different times due to factors such as forgetting and the evolving nature of their musical tastes ~\cite{jawaheer2014modeling, amatriain2009rate, sanna2021next}. Consequently, a song that a user played repeatedly months ago may no longer reflect their current preferences. 

These modulating factors suggest that certain variables may serve as indicators of more reliable feedback. For example, a sufficient number of interactions with a song may be necessary for the~\ac{MEE} to affect preference formation adequately. Additionally, ensuring that recent interactions have occurred within a relatively short period could mitigate the potential confounding effects of temporal factors such as forgetting or changes in personal taste.

In this section, we analyze aggregated listening feedback across parameters that capture both exposure and time factors, aiming to uncover patterns of uncertainty in user interactions. Specifically, we focus on estimating uncertainty from implicit feedback, accounting for the key factors mentioned above. Our study leverages interaction logs from Deezer, a globally available music streaming platform, with a primary focus on inter-user consistency, surfacing global trends in behavioral patterns. This approach allows us to identify variations in listening patterns across users, providing insights into the consistency and reliability of user behavior.

\subsection {Data}
\label{sec:data}

We collected users' listening histories from January 2022 to May 2023, excluding tracks played before July 2022 to ensure we capture the initial interactions of the users with the remaining items. We filtered out users that interacted with fewer than 20 unique tracks and tracks that were interacted by fewer than 100 unique listeners. We define a \ac{LE} as the streaming of a track for at least 30 seconds, which is a threshold largely used in the industry for remuneration purposes. Since the distribution of  the number of repetitions per user-track pair is highly skewed, with a median of 1 and a 99th percentile of 57. We excluded 1\% of pairs where users repeated tracks more than 57 times. To represent varying interaction levels, we sampled pairs of users and items based on their total number of repetitions and kept all their interactions.

The final dataset includes approximately 11 million interactions, 40,600 unique users, and 12,500 unique tracks. Each record comprises a user identifier, item identifier, timestamp, and a binary variable $L$ indicating if the \ac{LE} is positive ($L=1$) or negative ($L=0$), corresponding to a skip. We make the data available in our GitHub repository (see Section~\ref{sec:results}).

\subsection{Quantifying  Uncertainty in Repeated Consumption}

The literature on~\ac{UQ} in machine learning distinguishes between two primary types of uncertainty: \ac{AU}, which arises from inherent variability in the data, such as class overlap or noise in the observed phenomenon; and \ac{EU}, which stems from the model's lack of knowledge, typically due to insufficient or unrepresentative data. Together, these form the total uncertainty (TU), which can be expressed as the sum of both components: $TU = AU + EU$~\cite{abdar2021review}.

Uncertainties can be quantified as probabilities and interpreted either through the frequentist lens, where probabilities represent frequencies over repeated trials, or from a Bayesian perspective, where they reflect degrees of belief. Bayesian approaches are particularly effective in deriving uncertainty estimates~\cite{gregory2005bayesian, abdar2021review, contreras2018bayesian, johnson2022bayes}. In Bayesian terms, a model's parameters $\theta$ are seen as hypothesis or explanations about the world, given a prior over the parameter distributions $P(\theta)$, some data $D = \{x_i\}_{i = 1}^N$ as evidence, a likelihood function $P(D|\theta)$ and a normalization factor $P(D)$ (the marginal likelihood obtained by integrating over the entire parameter space), Bayes’ rule is applied: 

\begin{equation*}
P(\theta|D) = \frac{P(D|\theta) P(\theta)}{P(D)} \quad  \text{read:} \quad Posterior = \frac{Likelihood \times Prior}{Evidence} 
\end{equation*}



for obtaining the posterior distribution $P(\theta|D)$ that  serves as the degree of belief about the hypothesis $\theta$ given the evidenced data $D$. 

In this work, we focus on modeling user listening behavior derived from \ac{LE}s, where $L$ is a binary variable indicating whether a user listens to a song (1) or skips it (0). Aleatoric uncertainty appears when user behavior shows significant variability—i.e., there is no clear distinction between listening and skipping, resulting in high variance in $L$, e.g., when $P(L=1)$ fluctuates around 0.5, reflecting nearly random behavior. Epistemic uncertainty arises when the data fails to fully capture user preferences, leading to incomplete knowledge about behavior.

To quantify aleatoric uncertainty, we estimate the expected value of $L$, which corresponds to $P(L=1)$, given its binary nature. We model listening behavior using a beta distribution as prior, as it is well-suited to representing probabilities of binary outcomes. The beta distribution is particularly useful here as it is a conjugate prior for the binomial distribution, meaning it provides a natural framework for updating beliefs about probabilities given binary outcomes such as listens or skips.

The beta distribution, denoted $Beta(\alpha, \beta)$, is a continuous distribution defined on the interval $[0,1]$ and is parameterized by two shape parameters, $\alpha$ and $\beta$. The \ac{pdf} of the Beta distribution for a value $\pi$ on the interval (0,1) is: 

\begin{equation*}
g(\pi|\alpha,\beta)= \frac{\pi^{\alpha-1}(1-\pi)^{\beta-1}}{B(\alpha, \beta)},\label{eq:BetaDensity}
\end{equation*}

where $B(\alpha,\beta)$ is a normalization constant given by $\frac{\Gamma(\alpha + \beta)}{\Gamma(\alpha)\Gamma(\beta)}$, and $\Gamma(x)$ is the Gamma function. For practical purposes, instead of explicitly computing this normalization constant, we can simplify the posterior construction by leveraging the fact that the posterior \ac{pdf} is proportional to the product of the prior \ac{pdf} and the likelihood function \cite{johnson2022bayes}. Suppose the prior $Beta(\alpha,\beta)$,
and we observe a binomial sample $Y$ with $n$ trials and $y$ successes. Then, since the posterior is proportional to the prior times the likelihood, the posterior \ac{pdf} will be proportional to $g(\pi|y) \propto \pi^{y + \alpha -1}(1 -\pi)^{n-y+b-1} $ 
 which is the beta distribution with updated parameters $\alpha' = \alpha +y$ and $\beta' = \beta + n - y$, for more details we recommend~\cite{bolstad2016introduction}.

\subsubsection{Uncertainty over play counts}
Our first investigated dimension is the number of play counts, or exposures. For each interaction of a user $u$ and item $v$ in our data, we compute the play count (number of \ac{LE}s) the user has accumulated with that track up to that point in time, denoted $playcount(u,v,i)$, where $i$ is the $i$-th interaction of $u$ and $v$. Here we make a distinction between interaction and ~\ac{LE}. An interaction $i$ refers to either $L_i = 1$, if the corresponding listening time is $\geq 30s$ or $L_i = 0$ otherwise, which corresponds to a skip. Note that at interaction $i$ for a pair of user-item, $playcount(u,v,i) = \sum_{j = 1}^{i-1} L_j$. In the case of multiple skips, the $playcount$ does not change, resulting in multiple interactions with the same value of $playcount$. 

We assume that user interactions at a given level of $playcount$ are independent of one another. To ensure this independence, we retain only the first interaction of each user-item pair following the previous listening event \ac{LE}. We define $\pi_k$ as the listening tendency, conditioned on the fact that a user has listened to the item $k$ times, i.e., $mean(\pi_k) = E(L|playcount = k)$, since $L$ is binary. We note that user-item pairs whose single interaction was a \ac{LE} are excluded from this analysis (though they are included in later tests). These pairs are disproportionately common in our dataset due to the sampling process, which retains only sequences of interactions with at least one repetition. Including these pairs would misrepresent repetitive patterns, as such pairs are absent from other $playcount$ levels. Therefore we assume that the first interactions of all users at the $k$-th $playcount$ are independent of one another. 

When selecting the prior for our beta distribution, we centered it around $0.5$ to reflect random chance for $L$. The parameters $\alpha$ and $\beta$ determine the informativeness of the prior and the relative weight given to the actual data. We set $\alpha = 5$ and $\beta = 5$ which corresponds to a weak prior, as we are letting the data drive the posterior. For every $playcount$ level we update $\alpha$ with the number of positive interactions and $\beta$ with the number of skips as described above.

\begin{figure*}[ht]
    \centering
    \includegraphics[width=0.8\textwidth]
    {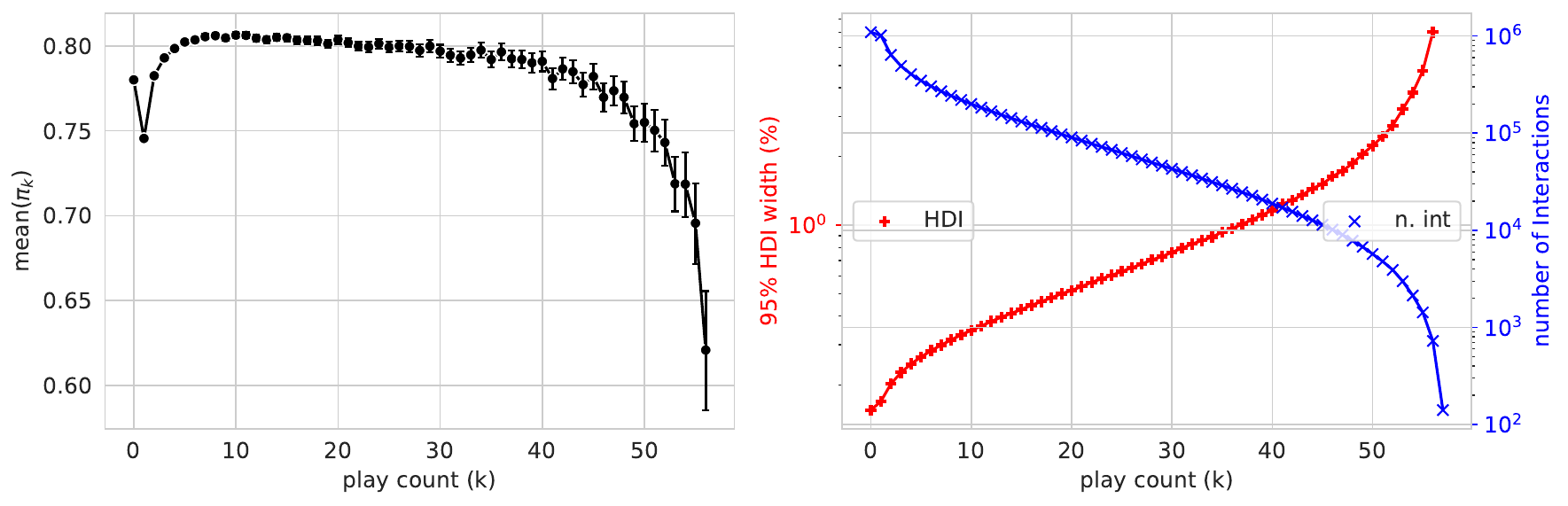}
    \caption{Left: the evolution of the expected value of the posterior distributions per level of $playcount$. Right: in red the evolution of the 95\% \ac{HDI} of the posteriors per $playcount$ and in blue the number of interactions per $playcount$.}
\label{fig:expected_playcount}
\end{figure*}

On the left of Figure~\ref{fig:expected_playcount}, we present the mean $\pi_k$ of the posterior distribution for each \textit{playcount} level, alongside the corresponding 0.95~\ac{HDI}, that is, the interval containing 95\% of the posterior probability mass. On the right side of Figure~\ref{fig:expected_playcount} we depict in blue the number of interactions per level of $playcount$ and in red the evolution of the 0.95~\ac{HDI}.

 The observed difference in the mean value of $\pi_k$ for $k = 0$ and $k > 0$ likely derives from the users' interest in listening to a song the first time, the behavior after the first listen then being more representative of preferences. The curve representing the mean $\pi_k$ value shows the characteristic inverted-U shape of the~\ac{MEE}: as users are more exposed to the items, their understanding of the music also augments, resulting in an increase in liking, however at one point interest peaks (at around $k=10$) and every subsequent exposure results in satiation~\cite{zajonc1968attitudinal, sguerra2022discovery}. The decrease in the mean over repetitions indicates that, as $playcount$ increases, some users remain engaged and continue listening to the song, while others become disinterested and cease listening. This overlap of behaviors is what is being measured by our posteriors, indicating changes in~\ac{AU}. On the other hand, the more observations for a given level of $playcount$, the more narrow the~\ac{HDI}, as shown. The higher the level of $playcount$, the fewer users remain interacting with the items,  which in turn increases the width of~\ac{HDI} indicating changes in~\ac{EU}.

\subsubsection{Uncertainty over recency} 

The previous section analyzed listening patterns based on user groups with similar behavior in terms of play count. However, it overlooked the time intervals between these interactions. This is problematic, as longer intervals between repetitions may signal boredom~\cite{benson2016modeling}, and could also relate to memory decay, potentially alleviating satiation~\cite{reiter2021predicting, hu2011nextone}. Moreover, recency is known to be one of the strongest predictors of repeated consumption~\cite{anderson2014dynamics}.

Hence, for each item $v$ interacted by user $u$, we compute the time between an interaction with the last~\ac{LE} of $v$ by $u$ (i.e., the relistening recency). However, unlike the discrete nature of play count, time is a continuous variable. Furthermore, the number of interactions in the dataset follows a power-law decay over increasing time differences, which is a well-documented pattern in repeated consumption research~\cite{kowald2017temporal}. 

\begin{figure*}
    \centering
    \includegraphics[width=0.8\textwidth]
    {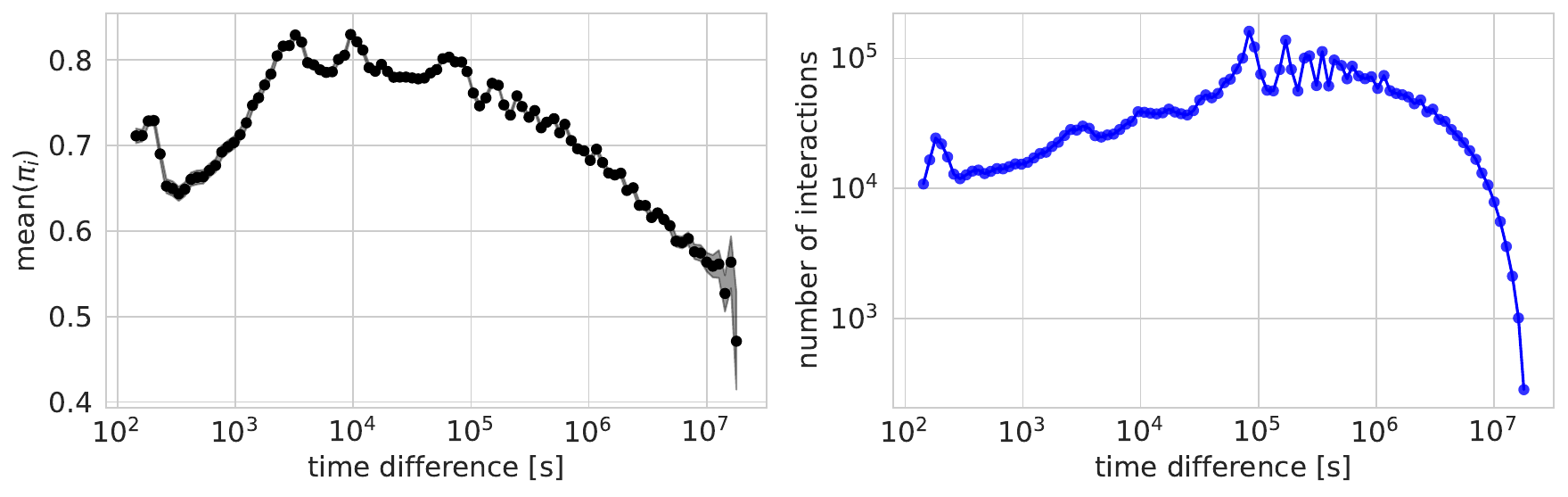}
    \caption{Left: the evolution of the expected value of the posteriors obtained for the recency bins together with a shaded 95\% \ac{HDI}. Right: the number of interactions per recency bin. }
\label{fig:expected_time}
\end{figure*}

To aggregate users' interactions into meaningful and representative groups, we discretize the relistening recency on a logarithmic scale to capture the decaying engagement patterns. We model the listening behavior within each recency bin using a beta distribution, with a weak prior set with $\alpha = 5$ and $\beta = 5$. The observed interactions in each bin serve as binomial evidence, allowing us to compute the posterior distributions, as described above. The left of Figure~\ref{fig:expected_time} illustrates the evolution of the mean $\pi_i$ together with the a shaded area representing 95\% of the~\ac{HDI} obtained from the posterior for each $i$-th bin, plotted against the mean recency value per bin. The right side of Figure~\ref{fig:expected_time} depicts the number of interactions at each of the bins. Here we consider the behavior for time differences superior to 134s, which is the lower end of the track duration in our data.

The resulting curve reveals peaks associated with periodic consumption behaviors. For instance, the first peak appears around the average duration of a song in the dataset. The high $\pi_i$ value here indicates that, when a song replayed in close succession, users had the tendency to not skip it, which probably derives from periods of high interest when users are listening to a track in loop.  Other periodic peaks, such as the one observed around the 24-hour mark ( \(\sim \)$10^{4.9}$), align with well-documented trends in repeated consumption research~\cite{kowald2017temporal}. It is worth noting that the highest $\pi_i$ values do not necessarily coincide with areas of greatest interaction frequency, as illustrated on the right side of Figure~\ref{fig:expected_time}.

\section{From uncertainty to reliability}
\label{sec:reliability}
The previous section focused on estimating uncertainty in listening behavior across two key dimensions: play count and time interval. We demonstrated that distinct patterns of intra-user uncertainty emerge solely from repeated consumption, without the need to incorporate content information or personalized taste metrics. These uncertainty estimates are driven by recurring behavioral patterns—such as relistening to the same track after 24 hours, which reflects a high probability of positive interaction ($L = 1$), or low~\ac{AU}. We propose that these low-uncertainty patterns provide more reliable feedback, making them critical for robust user preference modeling.

In contrast to the previous analysis, where we examined each dimension independently, here we jointly quantify uncertainty across both dimensions. This enables us to capture behaviors that depend simultaneously on recency and exposure. However, discretizing the dataset along these two dimensions introduces sparsity, reducing the amount of available evidence for computing reliable posteriors. This sparsity can cause the posterior estimates to become unstable when driven by scarce interaction data. To mitigate this, we can either adjust the number of bins used for discretizing recency or modify the prior assumptions in the beta distribution. Specifically, setting higher values for the prior parameters $\alpha$ and $\beta$ reduces the influence of the likelihood on the posterior, stabilizing the estimates.

\begin{figure*}%
    \centering
    \subfloat[\centering]{{\includegraphics[width=0.44\textwidth]{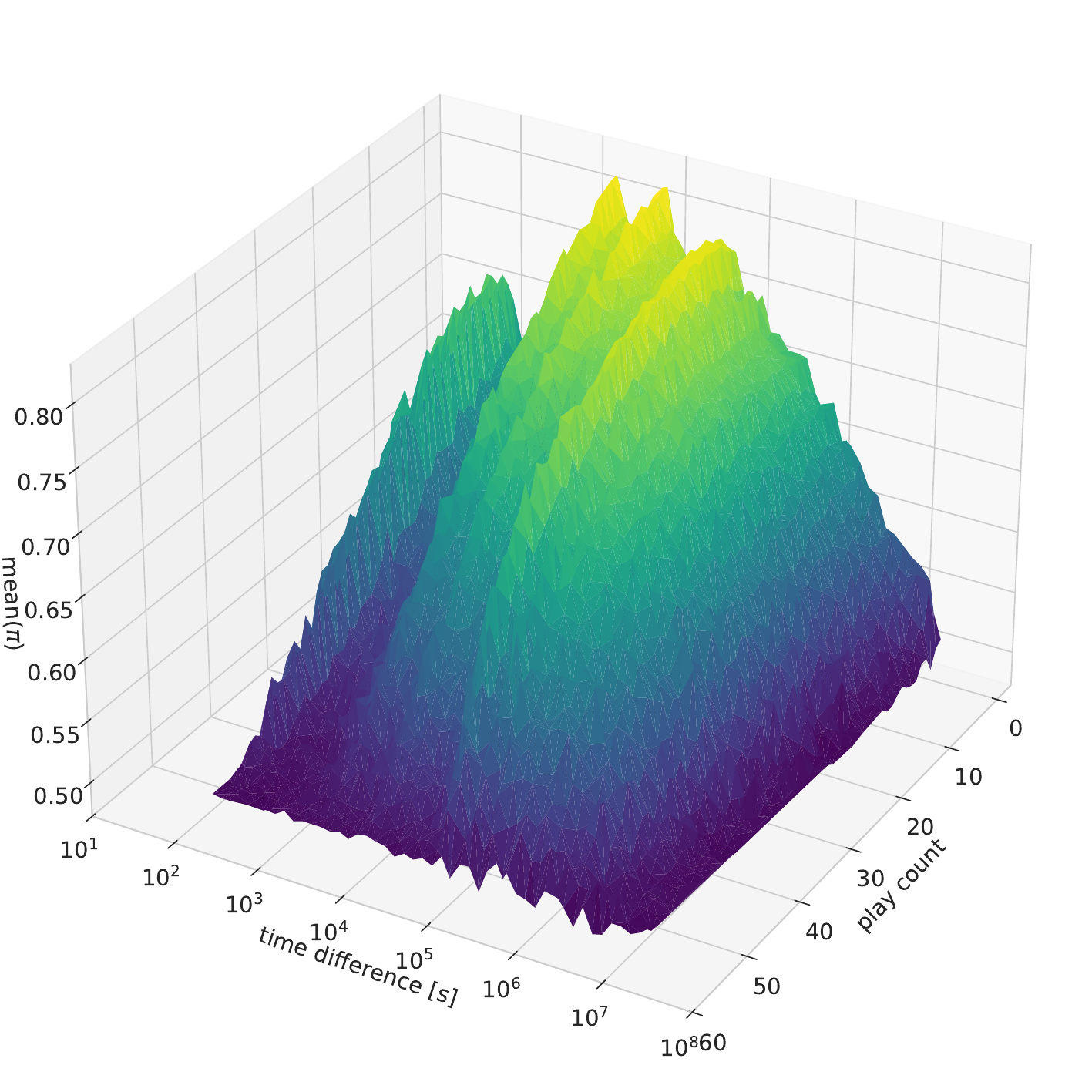} }}%
    \subfloat[\centering]{{\includegraphics[width=0.36\textwidth]{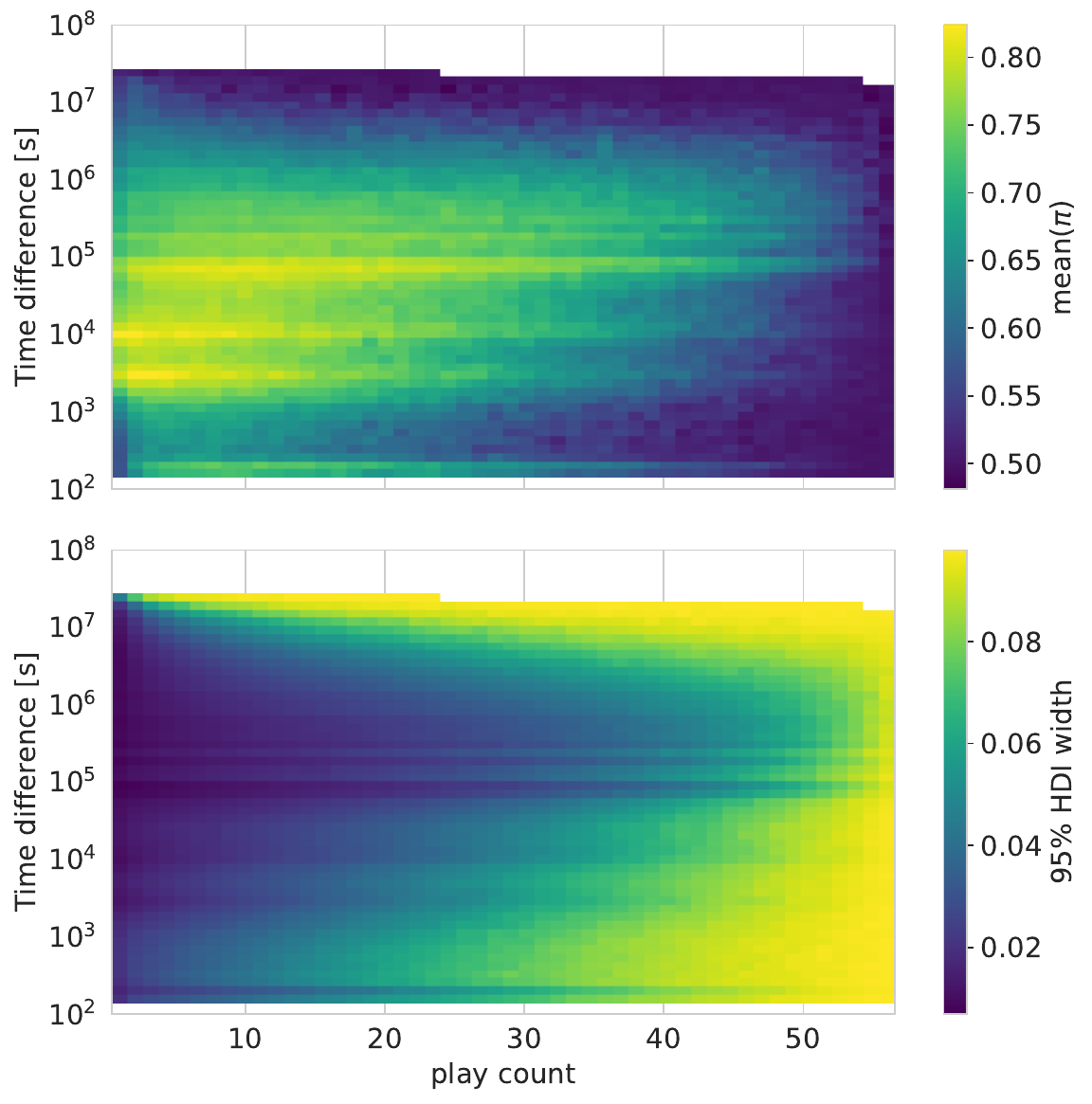} }}%
    \caption{(a) a 3D representation of the average posterior distribution over recency and $playcount$ combinations with a prior $Beta(200,200)$; (b) a 2D projection of the mean posteriors and the corresponding 95\% \ac{HDI}.}
    \label{fig:plane}%
\end{figure*}

Figure~\ref{fig:plane} illustrates this approach: the left panel shows a 3D plot of the mean posterior values obtained across 50 recency bins and varying play count levels, where we use a stronger prior with $\alpha = \beta = 200$. The right panel provides a 2D projection, highlighting both the mean posteriors and the corresponding 95\% \ac{HDI}. We observe that the peaks corresponding to the recurring patterns in Figure~\ref{fig:expected_time} remain visible, though they tend to be attenuated as the play count increases. When either $playcount$ and the time difference increases, the number of characterizing interactions become sparser, therefore the posterior is heavily guided by the prior resulting in $mean(\pi) = 0.5$.

These posteriors serve two primary purposes: (i) the mean of each posterior measures the degree of overlap, or lack thereof, between listening and skipping behavior, given the number of past interactions and listening recency; (ii) the $HDI$ indicates our confidence in the posterior mean. Together, these metrics reveal how polarized a user's interaction is for a given pair of $playcount(u,v,i)$ and $recency(u,v,i)$, as well as how confident we are in this measure.

Since the posteriors are derived from aggregated behavior, they capture general trends in consumption. For users whose listening behavior aligns with these patterns, their interactions will fall into regions characterized by both low $HDI$ and high $mean(\pi)$. In contrast, users whose behavior deviates from these norms—such as those who consistently listen to the same track even at high levels of $playcount$—will have interactions located in areas with high $HDI$ and low $mean(\pi)$, reflecting uncertainty or atypical consumption patterns. 

While individual preferences or habits may lead to deviations from these patterns, for the average listener, these computed posteriors can help detect noisy or anomalous interactions. Interactions that deviate from typical consumption may signal behavior driven by factors other than interest. Therefore, the belief derived from our posterior distributions can help identify interactions most strongly tied to consistent, positive engagement.

Given the derived posteriors, we can approximate the behavior over the entire space of $playcount$ and $recency$ using linear interpolation. This approach enables us to estimate both the mean posterior and the \ac{HDI} for any arbitrary combination of playcount and recency within the characterized space. Let $\hat{\pi}(u,v,i)$ denote the approximated mean posterior for user $u$ and item $v$ at the $i$-th interaction, where each interaction is characterized by specific values of playcount and recency. Similarly, the width of \ac{HDI} for any interaction can be approximated and will be noted as $\hat{HDI}(u,v,i)$.

These continuous approximations allow us to infer both the expected mean posterior and the \ac{HDI} width for any user interaction based on the values of playcount and recency, even for combinations not directly observed in the dataset. For interactions with playcount and recency values that fall outside the observed limits, we default to using the prior distribution to guide our estimations. Notably, this applies to the very first interaction, where recency is undefined.

\section{Experimental Analysis}
\label{sec:experiments}
We evaluate the effectiveness of our uncertainty measures by incorporating them as confidence weights for user implicit feedback. In this evaluation, we employ the popular Implicit Matrix Factorization (ALS) model~\cite{hu2008collaborative}, introduced at the beginning of Section~\ref{sec:related_work}, which remains a competitive benchmark for implicit feedback tasks~\cite{ferraro2020maximizing, huang2024going}. One key advantage of \ac{ALS} is its flexibility in assigning different weights to user-item interactions, making it a fair framework for testing our proposed approach.

Our focus here is on improving the weighting of implicit feedback rather than introducing a new recommendation model, and the use of a single model isolates the impact of our weighting strategy. While models like NeuWMF~\cite{hennekes2023weighted} also accommodate weighted interactions, they rely on the Adam optimizer, which can exhibit instability when handling data batches with highly varied weights~\cite{huang2024going} (resulting in the authors also opting for \ac{ALS} in a similar scenario).

Using the \ac{ALS} model as a basis, we compare the performance of our uncertainty measures as confidence scores against commonly used weighting methods for positive interactions. It is common to consider the confidence in the general form $c_{u,v} = \alpha w_{u,v}$, where $\alpha$ is an hyperparameter and several possible choices may be considered for the weight $w_{u,v}$. We note that in~\cite{hu2008collaborative}, the authors propose adding 1 as a base confidence for all items, whether interacted with or not. However, this choice densifies the sparse interaction matrix, which can negatively impact scalability and efficiency. Therefore, the ALS implementation we use does not adopt this approach. Instead, we assign a confidence of 0 to items the user has not interacted with. The different weighting schemes tested in this evaluation include:

\begin{itemize}
    \item \textbf{linear}: $r_{u,v}$; where $r_{u,v}$ is the sum of positive interactions, as originally proposed by Hu et al.~\cite{hu2008collaborative}.
    
    \item \textbf{logarithmic}: $\log(1 + \frac{r_{u,v}}{\epsilon})$; also proposed by Hu et al.~\cite{hu2008collaborative}.
    
    \item \textbf{log popularity}: $c_{u,v} = \log(1 + \frac{r_{u,v}}{\tilde{r_v} \epsilon})$; where $\tilde{r_v}$ is the average number of positive interactions per item, as proposed in~\cite{hennekes2023weighted}.
     
    
    \item \textbf{sum of expected posteriors}: $\sum_{i} \mathbb{E}[\hat{\pi}(u,v,i)]$, the sum of the expected values of the approximated posteriors for all the interactions between user $u$ and item $v$, reflecting the aggregated reliability of the interactions.

    \item \textbf{log-sum of expected posteriors}: $\log( 1 +\sum_{i} \mathbb{E}[\hat{\pi}(u,v,i)])$, this transformation compresses larger values, preventing users with a high number of repetitions from disproportionately influencing the model's training. 

    
    \item \textbf{sum of confidences}: $\sum_{i} \frac{\mathbb{E}[\hat{\pi}(u,v,i)]}{ c + \hat{HDI}(u,v,i)}$, the sum of the confidence values for all interactions between user $u$ and item $v$. Here, confidence is defined as the ratio between the approximated expected value of the posterior and the width (the difference between the upper and lower bounds) of the approximated \ac{HDI}, denoted $\hat{HDI}(u,v,i)$. The term $c$ is a hyperparameter, serving as a cutoff to prevent very low $\hat{HDI}(u,v,i)$ from overly influencing the confidence.
\end{itemize}


\subsection{Evaluation Datasets}
We evaluate the weighting schemes described above by training~\ac{ALS} using two large-scale datasets from the music~domain:

\begin{itemize}
\item Last.fm \cite{schedl2016lfm}: This public dataset consists of over a billion time-stamped listening events from 
120k users of the music website Last.fm, encompassing 3M songs. 
We have filtered this dataset to include only the most recent year of consumption history as well as the tracks that appear for the first time in this period, ensuring that these interactions represent the user's first engagement with an item. Additionally, we filter out users who did not interact with at least 20 unique items and items that were not interacted with by at least 100 users.

\item Deezer: Our proprietary dataset described in Section~\ref{sec:data}.
\end{itemize}

Since the Last.fm dataset does not include listening time information, we derive $LE$s based on the time interval between subsequent interactions by the same user. An interaction is considered positive if it lasts more than 30 seconds; otherwise, it is treated as a skip.

Both datasets are time-split into training (first 70\% of the total time window) validation (from 70\% to 85\%) and test (the rest of the data). We further filter iteratively until every user in the train set has interacted with the minimum of 10 items and to ensure that each user has 2 unique items in both validation and test that they have not previously interacted with and that are present in the training data. We present the statistics of the resulting training sets in Table~\ref{table:datasets}.    

\begin{table}
\centering
\resizebox{1\columnwidth}{!}{
\begin{tabular}{ccccccc}
\hline
   training set & \#users &  \#items & \#int & median(\#rep) & mean(\#rep)  & max(\#rep) \\
\hline
Deezer & 20.8k  & 9.6k & 3.4M & 1 & 3.72 & 57\\

Last.fm & 4.1k  & 7.6k & 1.2M  & 1 & 3.45 & 73\\
\hline
\end{tabular}}
\caption{Statistics on the training sets. Both the validation and test sets consist of 2 unique items per user in the training set. We ensure the items were not previously interacted with by these users and that they appear in the training set.}
\label{table:datasets}
\end{table}

\subsection{Results and Discussion}
\label{sec:results}
 We examine the effectiveness of the weighting schemes detailed above on the two datasets through the recommendation metrics: Recall@K and \ac{NDCG}@K. We run ten experiments to assess variations in the performance due to initialization. The resulting scores are shown in Table~\ref{table: evaluation}. We optimize the weight parameters $\alpha$ in \{0.1, 0.5, 1, 1.5, 2, 10, 40, 100\}, $\epsilon$ (for the logarithmic weights) in \{0.1, 0.5, 0.8, 1, 1.5\} and the number of latent dimensions \{16, 32, 64\} on the validation set. The uncertainty measures were obtained from the train set only. For the Deezer dataset, we use 32 latent dimensions, a $Beta(500,500)$ prior, and 100 recency bins. For Last.fm, we use 64 dimensions, a $Beta(100,100)$ prior, and discretize recency into 5 bins. The code for computing uncertainty measures, running the experiments and the associated data are available on our GitHub repository\footnote{\url{https://github.com/deezer/uncertainty_feedback}}.

\begin{table*}[t]
\centering
\resizebox{0.8\textwidth}{!}{
\begin{tabular}[t]{cccccc}
\hline
  Data  & Weight &  Recall@10(\%) & Recall@20(\%) & NDCG@10(\%) & NDCG@20(\%) \\
\hline
\multirow{6}{*}{Deezer} 
& linear & 2.385 $\pm$ 0.028 & 4.754 $\pm$ 0.047 & 1.289 $\pm$ 0.018 & 2.013 $\pm$ 0.011 \\
& logarithmic & 2.360 $\pm$ 0.031 & 4.693 $\pm$ 0.042 & 1.252 $\pm$ 0.012 & 1.966 $\pm$ 0.011 \\
& log popularity & 2.366 $\pm$ 0.029 & 4.720 $\pm$ 0.029 & 1.260 $\pm$ 0.017 & 1.980 $\pm$ 0.013 \\
& sum expected posteriors & 2.424 $\pm$ 0.032 & \textbf{4.835*$\pm$ 0.081} & 1.330 $\pm$ 0.015 & \textbf{2.069**$\pm$ 0.029} \\
& log-sum expected posteriors & 2.366 $\pm$ 0.029 & 4.720 $\pm$ 0.029 & 1.260 $\pm$ 0.017 & 1.980 $\pm$ 0.013 \\
& sum confidences & \textbf{2.467**$\pm$ 0.035} & 4.805 $\pm$ 0.052 & \textbf{1.349**$\pm$ 0.022} & 2.065 $\pm$ 0.023 \\

\hline
\multirow{6}{*}{Last.fm} 
& linear & 3.304 $\pm$ 0.131 & 5.721 $\pm$ 0.175 & 1.927 $\pm$	0.083 & 2.673 $\pm$ 0.090 \\
& logarithmic	& 3.434 $\pm$ 0.079 & 5.723 $\pm$ 0.100 & 1.990 $\pm$ 0.057 & 2.695 $\pm$ 0.050 \\
& log popularity & 3.252 $\pm$ 0.086 & 5.615 $\pm$ 0.100 & 1.916 $\pm$ 0.050 & 2.644 $\pm$ 0.048 \\
& sum expected posteriors & 3.435 $\pm$	0.121 & 5.801 $\pm$ 0.084 & 2.008 $\pm$ 0.081 & 2.736 $\pm$ 0.055 \\
& log-sum expected posteriors & \textbf{3.51**$\pm$ 0.063 }& \textbf{5.801**$\pm$ 0.084} & \textbf{2.049**$\pm$ 0.048} & \textbf{2.757**$\pm$ 0.049} \\
& sum confidences & 3.330 $\pm$ 0.126&	5.722 $\pm$	0.158& 1.944 $\pm$ 0.063 & 2.682 $\pm$ 0.068 \\

\hline
\end{tabular}}
\caption{Experimental results (values reported as percentages). We conducted hypothesis testing using a t-test to compare the performance of 20 runs of the best-scoring model against the best-performing baseline for each dataset. Statistical significance is indicated as follows: $^*$ :$p<0.05$ and $^{**}$: $ p< 0.001$.}
\label{table: evaluation}
\end{table*}

As mentioned earlier, for users who do not repeat a song more than once, a common occurrence in our dataset as seen in Table~\ref{table:datasets}, we rely solely on the prior to estimate uncertainty. However, the subset of users who exhibit repeated listening behavior provides enough signal to enhance the training of \ac{ALS}, leading to improved user-item representations and better overall performance. This suggests that incorporating uncertainty patterns not only produces more robust confidence scores but also helps identify more reliable user feedback, ultimately driving better recommendations. 

Lastly, we highlight the performance differences between the two datasets. The uncertainty and repetitive patterns when fitting Beta distributions on Last.fm data differ from those in our proprietary dataset and appear more stable, exhibiting less variability. This discrepancy is partly due to the absence of listening time in the Last.fm dataset, which reduces both the number of skips and the precision of these interaction signals. The higher expected values of the posteriors in this dataset partly explain why the log-sum of expected posteriors performs better. Despite these differences, our approach demonstrates flexibility in adapting to dataset characteristics, consistently improving recommendation performance with dataset-specific weighting schemes.

\section{Conclusion and Future Work}
\label{sec:conclusion}
In this work, we conducted an in-depth exploration of uncertainty patterns in repeated user interactions. Specifically, we derived measures for \ac{AU} (through the posterior mean) and \ac{EU} (through the \ac{HDI}) from the posterior distribution of listening/skipping tendencies and demonstrated that these patterns align with well-known phenomena, such as the \ac{MEE} and the periodicity of human behavior. By leveraging quantified uncertainty as a measure of reliability in a recommendation task, our results show that modeling uncertainty for users with repeated interactions yields better performance than traditional weighting schemes. These findings underscore the importance of integrating uncertainty into feedback models to produce more accurate and reliable user preference representations. 

Beyond \ac{ALS}, we expect our approach to be transferable to more recent methods, as uncertainty modeling is a central issue in recommender systems working with implicit data. For instance, it could be adapted to loss strategies similar to those in \cite{wang2021denoising}, where data samples with high levels of uncertainty could be reweighted or truncated during the training of neural models. Additionally, incorporating uncertainty can help reweight repeated interactions, as demonstrated in the weighting schemes presented above, improving repeat-aware models such as ~\cite{tran2024transformers}.

The framework presented here opens avenues for further research. For instance, we focused on consistency measures for behavior characterized by repetitive patterns, but additional variables could be introduced into our posterior models. Since music preferences are highly contextual and exhibit varying consumption patterns \cite{marey2024modeling}, incorporating the context into our uncertainty models could account for differences in listening habits, such as seasonal music (e.g., Christmas songs) or background music used for work or study, which may have distinct repetition patterns. Other variables, such as user demographics, mainstreamness or music genre preferences, could further refine these models. With richer data, we could derive more nuanced confidence measures beyond the simple 2D model discussed here.

Although our study focused on music streaming, the proposed approach is broadly applicable to other domains. For instance, in social media platforms, where the short duration video format is predominant, repetition often occurs at a higher level of abstraction, such as revisiting the same creators or content types. This behavior leads to phenomena like satiation or the \ac{MEE}, with likely distinct uncertainty patterns than the ones presented here. The simplicity of our model allows it to be adapted to the specificities of such domains, offering a valuable framework for understanding user interactions in contexts where repetition plays a role.

Additionally, as outlined in Section~\ref{sec:reliability}, our uncertainty measures reflect average consumption patterns. Consequently, when deriving uncertainty for individual users, their uncertainty distribution can serve as an indicator of how ``typical'' their consumption behavior is. By clustering users based on these uncertainty measures, we can hierarchically model uncertainty within different user subgroups. This enables for more precise and tailored uncertainty estimates for various user communities, while also helping practitioners better understand them.

\bibliographystyle{ACM-Reference-Format}

\end{document}